\newif\ifconfver
\confverfalse      %declaring conference version false
\ifconfver
	\documentclass{article}
	\usepackage{spconf,amsmath,graphicx}
	\ninept
\else
	\documentclass[11pt]{article}
	\usepackage{fullpage}
\fi

\usepackage{calc,amsfonts,amssymb,amsmath,bm,url,color,theorem,graphicx,cite}
\usepackage{algorithm}
\usepackage{algorithmic}
\usepackage{soul}
\usepackage{enumerate}

\definecolor{orange}{RGB}{255,107,0}

\newtheorem{Fact}{Fact}
\newtheorem{Asm}{Assumption}
\newtheorem{Lemma}{Lemma}
\newtheorem{Prop}{Proposition}
\newtheorem{Theorem}{Theorem}

\theorembodyfont{\rmfamily}

\newcommand\bx{\ensuremath{{\bm x}}}
\newcommand\by{\ensuremath{{\bm y}}}
\newcommand\bG{\ensuremath{{\bm G}}}

\newcommand\bR{\ensuremath{{\bm R}}}

\newcommand\bX{\ensuremath{{\bm X}}}

\newcommand\ba{\ensuremath{{\bm a}}}

\newcommand\bA{\ensuremath{{\bm A}}}

\newcommand\bg{\ensuremath{{\bm g}}}
\newcommand\bB{\ensuremath{{\bm B}}}

\newcommand\bPi{\ensuremath{{\bm \Pi}}}

\newcommand\bF{\ensuremath{{\bm F}}}

\newcommand\bYh{\ensuremath{{\bm Y}_{\rm H}}}
\newcommand\bYm{\ensuremath{{\bm Y}_{\rm M}}}

\newcommand\bymi{\ensuremath{{\bm y}_{{\rm M},i}}}
\newcommand\byhi{\ensuremath{{\bm y}_{{\rm H},i}}}

\newcommand\Lh{\ensuremath{L_{\rm H}}}
\newcommand\Mm{\ensuremath{M_{\rm M}}}

\newcommand\bY{\ensuremath{{\bm Y}}}

\newcommand\bs{\ensuremath{{\bm s}}}
\newcommand\bS{\ensuremath{{\bm S}}}

\newcommand{\Rbb}{\mathbb{R}}

\newcommand{\setA}{\mathcal{A}}

\newcommand{\setU}{\mathcal{U}}

\newcommand{\setS}{\mathcal{S}}

\newcommand{\setK}{\mathcal{K}}
\newcommand{\setL}{\mathcal{L}}
\newcommand{\setJ}{\mathcal{J}}

\newcommand{\bzero}{{\bm 0}}
\newcommand{\bone}{{\bm 1}}
\newcommand{\bI}{{\bm I}}

\newcommand{\supp}{{\rm supp}}

\newcommand{\sigmax}{\sigma_{\rm max}}
\newcommand{\sigmin}{\sigma_{\rm min}}

\newcommand{\setI}{\mathcal{I}}

\newcommand{\paperabstract}{
Coupled structured matrix factorization (CoSMF) for hyperspectral super-resolution (HSR) has recently drawn significant interest in hyperspectral imaging for remote sensing.
Presently there is very few work that studies the theoretical recovery guarantees of CoSMF.
This paper makes one such endeavor by considering the CoSMF formulation by Wei {\em et al.}, which, simply speaking, is similar to coupled non-negative matrix factorization.
Assuming no noise,
we show sufficient conditions under which the globably optimal solution to the CoSMF problem is guaranteed to deliver certain recovery accuracies.
Our analysis suggests that sparsity and the pure-pixel (or separability) condition  play a   hidden role in enabling CoSMF to achieve some good recovery characteristics.
}

\title{Is There Any Recovery Guarantee with Coupled Structured Matrix Factorization for Hyperspectral Super-Resolution?}
\ifconfver
	\name{Huikang Liu, Ruiyuan Wu and Wing-Kin Ma \thanks{This research is supported by project \#MMT-8115059 of the Shun Hing Institute of Advanced Engineering, The Chinese University of Hong Kong.}}
	\address{Department of Electronic Engineering, The Chinese University of Hong Kong, Hong Kong SAR of China}
\else
	\makeatletter
	\def\thanks#1{\protected@xdef\@thanks{\@thanks
			\protect\footnotetext{#1}}}
	\makeatother

	\author{Huikang Liu, Ruiyuan Wu and Wing-Kin Ma  \\ ~ \\
	Department of Electronic Engineering, 
	The Chinese University of Hong Kong, \\ Hong Kong SAR of China \\ ~ \\
	E-mails: hkliu2014@gmail.com, rywu@ee.cuhk.edu.hk, wkma@ee.cuhk.edu.hk
 		\thanks{This research is supported by project \#MMT-8115059 of the Shun Hing Institute of Advanced Engineering, The Chinese University of Hong Kong.}
	}
\fi

\begin{document}
%\ninept
%
\maketitle
\begin{abstract}
\paperabstract
\end{abstract}
\ifconfver
\fi

\section{Introduction}
\label{sec:intro}

Recently, in remote sensing, there has been a flurry of research in hyperspectral super-resolution (HSR).
The problem is to construct a super-resolution (SR) image---which possesses both high spectral and spatial resolutions---from a co-registered pair of multispectral (MS) and hyperspectral (HS) images \cite{loncan2015hyperspectral}.
The MS and HS images have limited spectral and spatial resolutions, respectively, owing to hardware constraints,
and the possibility of fusing the two to achieve super-resolution imaging is a very attractive idea.
HSR was empirically demonstrated to be possible in the pioneering research \cite{yokoya2012coupled,Kawakami}.
There, the approach is to formulate the problem as a coupled structured matrix factorization (CoSMF) problem.
For example, Yokoya {\em et al.} apply non-negative matrix factorization (NMF) in their famous coupled NMF (CNMF) algorithm \cite{yokoya2012coupled}.
Naturally one can also consider other structured matrix factorization (SMF) formulations, such as those involving sparsity, spatial smoothness, etc.
In fact, the majority of the current HSR research are focused on various CoSMF formulations and the subsequent algorithm designs.

Given the rapid development of HSR, and the empirical successes reported therein,
there is a strong motivation to understand whether CoSMF truly works---in theory.
More precisely, the question is about what are the conditions, both on the MS-HS sensor specifications and on the scene, such that CoSMF is provably guaranteed to yield certain recovery accuracies with the SR image.
Presently we see very few research on such theoretical direction,
in stark contrast to the numerous research on algorithm designs.
The only available work is our previous paper \cite{hsr_recovery_ssp2018}, which studies recovery guarantees under a decoupled SMF pathway.

In this paper we consider a CoSMF formulation similar to CNMF, 
namely, the one by  Wei {\em et al.} \cite{wei2016multiband}, in which the linear spectral mixture model structure is exploited.
%, namely, that by Wei {\em et al.} \cite{wei2016multiband} which led to the FUMI algorithm.
%It is a variant of CNMF,
%and it exploits the linear spectral mixture model structure.
The main contribution of this paper is to analyze the recovery accuracy of this CoSMF.
Assuming the noiseless case,
we identify sufficient conditions under which the recovery error of any globally optimal solution to the CoSMF problem can be bounded.
From there, we reveal insights on when CoSMF is theoretically guaranteed to perform well.
\ifconfver
Owing to space limitation, we are unable to include all the proofs in this paper.
The complete proofs can be found in the extended, and online-accessible, version of this paper \cite{temp}.
\else
\fi

Some notations in this paper are defined as follows.
Given a matrix $\bX \in \Rbb^{m \times n}$, $\bx^i \in \Rbb^{n}$ and $\bx_j \in \Rbb^{m}$ represent the $i$th row and $j$th column of $\bX$, respectively;
$\bX^\setI$ is a submatrix of $\bX$ obtained by keeping the rows of $\bX$ indicated by $\setI$;
similarly, $\bX_\setJ$ is a submatrix obtained by keeping the columns indicated by $\setJ$, and
$\bX_\setJ^\setI$ the rows indicated by $\setI$ and columns indicated by $\setJ$;
%$\Rbb_+$ is the set of all non-negative real numbers;
$\| \bx \|_0$ denotes the number of nonzero elements of $\bx$;
$\bX \geq \bzero$ and $\bX \leq \bone$ mean that $x_{ij} \geq 0$ and $x_{ij} \leq 1$ for all $i,j$, respectively.

\section{Model}
\label{sec:model}

\begin{figure*}[tbh]
	\centering
	\ifconfver
		\includegraphics[width=0.9\linewidth]{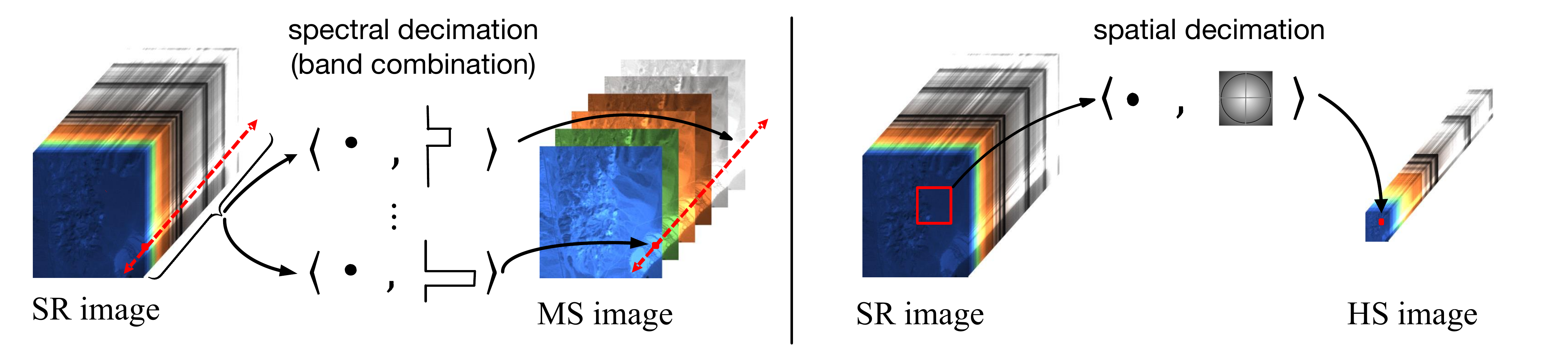}
	\else
		\includegraphics[width=0.99\linewidth]{decimation_model}
	\fi
	\caption{Illustration of the observation model.}
	\label{fig:illus}
\end{figure*}

The signal model of the HSR problem is described as follows.
Let $\bar{\bX} \in \Rbb^{M \times L}$ be the spectral-spatial matrix of the SR image we desire to obtain.
Here $M$ denotes the number of spectral bands, and $L$ the number of pixels.
The column $\bar{\bx}_i$ of $\bar{\bX}$ describes the spectral pixel at a specific spatial position indexed by $i$.
The SR image is incompletely observed by an MS sensor and an HS sensor.
Fig.~\ref{fig:illus} depicts how the relationship between the SR image and its MS-HS observations is modeled.
For the MS image, one MS spectral band is modeled as a linear combination of a number of contiguous SR spectral bands.
Specifically, the spectral pixels observed by the MS sensor are expressed as
%are modeled by
\begin{equation} \label{eq:bymi}
\bymi = \bF \bar{\bx}_i, \quad i=1,\ldots,L,
\end{equation}
where $\bF \in \Rbb^{\Mm \times M}$ describes the spectral decimation response;
$\Mm < M$ is the number of MS spectral bands.
%Specifically,  \eqref{eq:bymi} describes the MS observation process in which one MS spectral band is formed by combining a number of contiguous spectral bands.
Note that we assume no noise.
For the HS image, the HS pixels of each spectral band are modeled as a spatially blurred and down-sampled version of the SR counterpart.
Correspondingly, the spectral pixels observed by the HS sensor are given by
\begin{equation} \label{eq:byhi}
\byhi = \sum_{j \in \setL_i} \bar{\bx}_j g_{ji} = \bar{\bX}_{\setL_i} \bg_i,
\quad i=1,\ldots,\Lh,
\end{equation}
where $\setL_i \subset \{ 1,\ldots, L \}$ indicates a neighborhood of SR spectral pixels that have correspondence with the HS spectral pixel at position $i$;
$g_{ji}$ describes the spatial decimation response;
$\Lh < L$ is the number of HS pixels;
$\bg_i \in \Rbb^{|\setL_i|}$ is a vector obtained by concatenating the coefficients $\{ g_{ji} \}_{j \in \setL_i}$.
%The model \eqref{eq:byhi} represents one HS pixel as a spatially blurred and down-sampled version of some SR pixels.
The spatial decimation response vector $\bg_i$ satisfies
\begin{equation} \label{eq:g}
\bg_i > \bzero, \quad \bone^T \bg_i = 1;
\end{equation}
see, e.g., \cite{yokoya2012coupled}. 
Also we assume $\cup_{i=1}^{\Lh} \setL_i  = \{ 1,\dots, L\}$, and that 
one $\setL_i$ can have overlap with another.

The SR image is assumed to obey the linear spectral mixture model \cite{Jose12}, wherein every spectral pixel is posited as a linear combination of a number of distinct materials, or endmembers; that is,
\begin{equation} \label{eq:lmm}
\bar{\bx}_i = \textstyle \sum_{k=1}^N \bar{\ba}_k \bar{s}_{ki} =  \bar{\bA} \bar{\bs}_i, 
\end{equation}
where each $\bar{\ba}_k$ is the spectral signature of an endmember;
$\bar{s}_{ki} \geq 0$ describes the contribution, or abundance, of endmember $k$ in pixel $i$;
$\bar{\bA}= [~ \bar{\ba}_1,\ldots, \bar{\ba}_N ~]$;
$\bar{\bs}_i = [~ \bar{s}_{1,i},\ldots, \bar{s}_{N,i} ~]^T$;
$N$ is the number of endmembers.
The abundance vectors $\bar{\bs}_i$'s are typically assumed to lie in the unit simplex; i.e., 
\[
\bar{\bs}_i \in \setU^N := \{ \bs \in \Rbb^N \mid \bs \geq \bzero, ~ \bone^T \bs = 1  \}.
\]
The model order $N$ is much smaller than $M$ and $L$, but it can be greater than $\Mm$.
The endmember matrix $\bar{\bA}$ is non-negative by nature.
We may also assume $a_{ij} \leq 1$; this is because the MS-HS measurements are in the form of reflectance, with the range usually given by $[0,1]$.

For conciseness, we rewrite the model \eqref{eq:bymi}--\eqref{eq:lmm} as
%\begin{subequations} \label{eq:base_model}
%	\begin{align}
%	\bYm & = \bF \bar{\bX}, \\
%	\bYh & = \bar{\bX} \bG, \\
%	\bar{\bX} & = \bar{\bA} \bar{\bS},
%	\end{align}
%\end{subequations}
\begin{equation} \label{eq:base_model}
	\bYm  = \bF \bar{\bX}, \quad 
	\bYh  = \bar{\bX} \bG, \quad
	\bar{\bX}  = \bar{\bA} \bar{\bS},
\end{equation}
where $\bYm$ is a matrix obtained by concatenating the $\bymi$'s;
$\bYh$ and $\bar{\bS}$ are obtained by the same fashion;
$\bG \in \Rbb^{L \times \Lh}$ has its $(j,i)$th element given by $g_{ji}$ if $j \in \setL_i$, and by $0$ if $j \notin \setL_i$.
Also we denote
\[
\setU^{N \times L} = \{ \bS \in \Rbb^{N \times L} \mid \bs_i \in \setU^N, ~i=1,\ldots,L \}.
\]
%Note $\bar{\bS} \in \setU^{N \times L}$.
Recall $\bar{\bA} \in [0,1]^{M \times N}$, $\bar{\bS} \in \setU^{N \times L}$.

\section{Problem Statement}

The HSR problem is to recover the SR image $\bar{\bX}$ from the MS-HS image pair $(\bYm,\bYh)$.
Under the above introduced model, it is natural to consider the following CoSMF formulation
\begin{equation} \label{eq:CoSMF}
\min_{\bA \in \setA, \bS \in \setS} ~ \| \bYm- \bF \bA \bS \|_F^2 + \| \bYh-  \bA \bS \bG \|_F^2,
\end{equation}
where $\bF$ and $\bG$ are assumed to be known, which is done via calibration or estimation \cite{yokoya2012coupled,simoes2015convex};
\begin{equation} \label{eq:setAS}
\setA = [0,1]^{M \times N}, \quad \setS = \setU^{N \times L}.
\end{equation}
This CoSMF formulation was introduced in \cite{wei2016multiband}. 
It was inspired by the CNMF formulation \cite{yokoya2012coupled} which 
replaces \eqref{eq:setAS} by $\bA \geq \bzero, \bS \geq \bzero$.
%assumes $\setA = \Rbb_+^{M \times N}, \setS = \Rbb_+^{N \times L}$;
%here $\Rbb_+^{m \times n}$ denotes the set of all $m \times n$ non-negative matrices.
From the appearance of problem \eqref{eq:CoSMF}, one may have the impression that the aim is to identify $\bar{\bA}$ and $\bar{\bS}$, the true endmember and abundance matrices. 
This is not entirely true.
We are concerned with whether the solution $(\bA,\bS)$ leads to a reconstructed SR image $\bX = \bA \bS$ that is the same as, or close to, the true one $\bar{\bX}$.

\begin{figure}[hbt]
	\centering
	\ifconfver
		\includegraphics[width=0.75\linewidth]{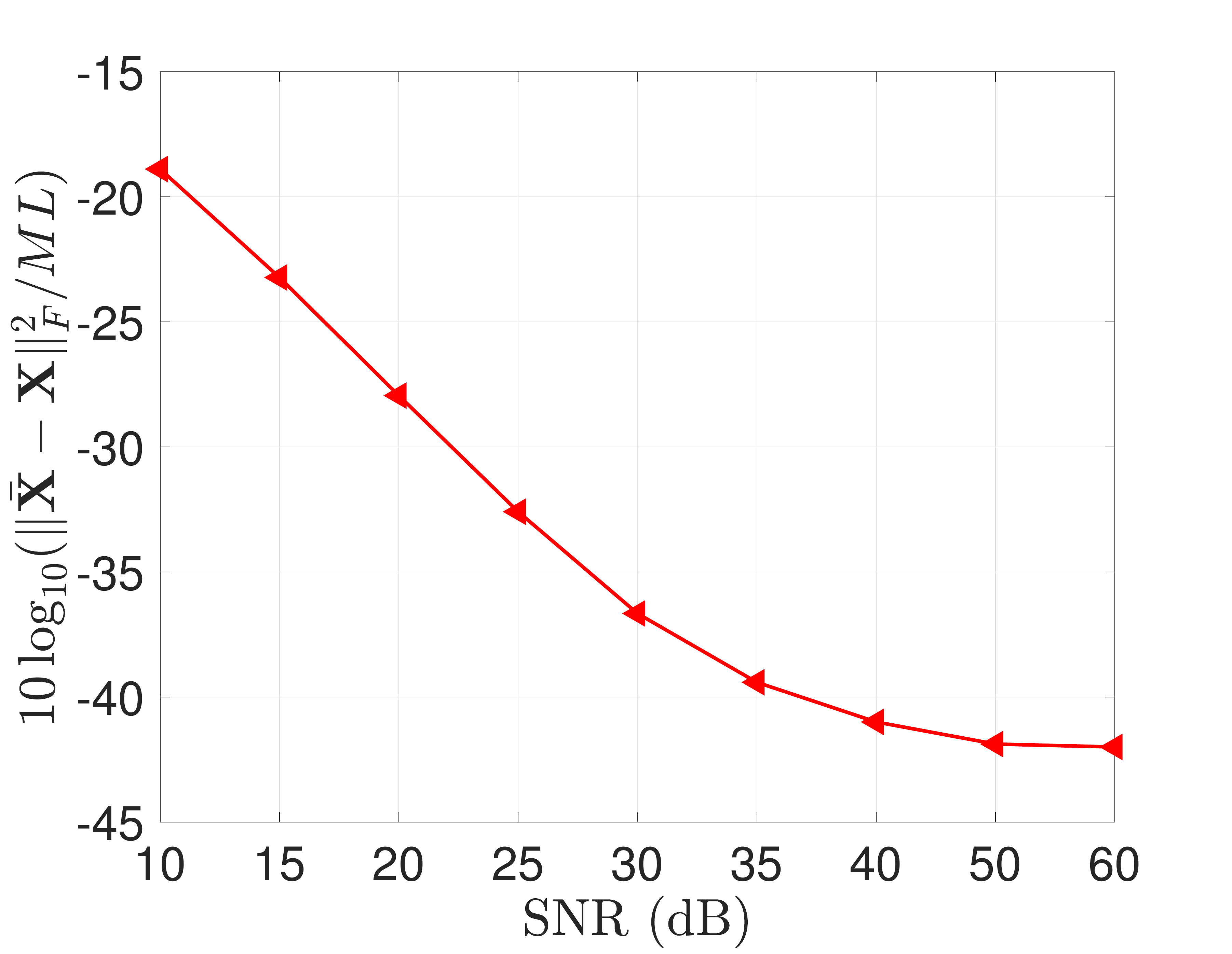}
	\else
		\includegraphics[width=0.6\linewidth]{semi_real}
	\fi
	\caption{MSE performance in a semi-real data experiment.}
	\label{fig:mse}
\end{figure}

Our interest in this paper is to show sufficient conditions under which the CoSMF problem \eqref{eq:CoSMF}--\eqref{eq:setAS} promises some kind of recovery guarantees with the true SR image.
Before we do so, we give the reader a glimpse of how CoSMF performs empirically.
Fig.~\ref{fig:mse} displays the average mean-squared error (MSE) of CoSMF in a semi-real data experiment.
Here, $\bar{\bX}$ is a real image, taken from the AVIRIS Indian Pine dataset \cite{vane1993airborne};
$\bYm$ and $\bYh$ are synthetically generated by \eqref{eq:base_model}, with noise added;
$\bF$ corresponds to the Landsat 4 TM MS sensor specification  \cite{chander2009summary};
$\bG$ corresponds to $11 \times 11$ Gaussian spreading (with variance $1.7^2$) and down-sampling of $4$;
we have $(M,\Mm,L,\Lh,N)=(178,6,120^2,30^2,30)$;
$100$ independent trials were tested;
the algorithm in \cite{ryan2018} is used to handle the CoSMF problem \eqref{eq:CoSMF}.
We see that the CoSMF provides reasonably good MSE performance under higher SNRs.

\section{The Main Result}

We first shed light onto the problem natures and put down assumptions along the way.
Consider an alternative representation of the model \eqref{eq:base_model}:
\begin{subequations}
\begin{align}
\bYm & = \bar{\bA}' \bar{\bS}, \label{eq:bYm_rewrite}  \\ 
\bYh & = \bar{\bA} \bar{\bS}', \label{eq:bYh_rewrite}
\end{align}
\end{subequations}
where 
\[
\bar{\bA}' = \bF \bar{\bA} \in \Rbb^{\Mm \times N}, \quad \bar{\bS}' = \bar{\bS} \bG \in \Rbb^{N \times \Lh}
\]
denote the spectrally-decimated true endmember matrix and the spatially-decimated true abundance matrix, respectively.
Particularly,
$\bar{\bS}'$ can be interpreted as the abundance matrix of the HS image $\bYh$.
It can be shown that $\bar{\bS}'$ lies in $\setU^{N \times \Lh}$ (use $\bar{\bs}_i'= \bS_{\setL_i} \bg_i$, \eqref{eq:g} and $\bS_{\setL_i} \in \setU^{N \times |\setL_i|}$).
We make the following assumption.
\begin{Asm} \label{Asm:1}
	The true endmember matrix $\bar{\bA}$ has full column rank.
	The spatially-decimated true abundance matrix $\bar{\bS}'$ has full row rank.
\end{Asm}
Assumption~\ref{Asm:1} is considered reasonable.
Physically it means that both the spectral signatures and HS abundance distributions
($\bar{\ba}_k$'s and $[ \bar{\bS}' ]^k$'s, respectively) of the various materials are distinctively different.
In fact, Assumption~\ref{Asm:1} is standard in the context of HS unmixing \cite{ma2013signal}.
Similarly we also want to assume that the spectrally-decimated signatures $\bar{\ba}_k'$'s of the various materials yield certain distinctiveness.
%without any of such assumption, 
%provable recovery may not be possible.
%we cannot do any recovery guarantee proof.
An easy way is to assume full column-rank $\bar{\bA}'$, but this is  generally not satisfied because $\bar{\bA}'$ can be fat: The number of MS spectral bands $\Mm$ is about $4$ to $8$ in the existing MS sensors, while the model order $N$ can easily be greater than $\Mm$.
For this reason we consider
\begin{equation} \label{eq:K}
K = {\rm krank}(\bar{\bA}'),
\end{equation}
where ${\rm krank}$ denotes the Kruskal rank,
and we use $K$ to quantify the distinctiveness of $\bar{\bA}'$.
Recall from the Kruskal rank definition that any collection of $K$ vectors in $\{ \bar{\ba}_1',\ldots, \bar{\ba}_N' \}$ is linearly independent.
Note $K \leq \min\{ N, \Mm \}$.

The abundances encountered in real life are usually sparse.
More precisely, an MS or HS pixel may be posited as a combination of a few materials.
Such sparse assumption has been used, argued to be reasonable, in sparse HS unmixing \cite{iordache2011sparse}.
This leads us to the following assumption.
\begin{Asm} \label{Asm:2}
Every column $\bar{\bs}_i'$ of the spatially-decimated true abundance matrix $\bar{\bS}'$ is $K$-sparse, i.e., $\| \bar{\bs}_i' \|_0 \leq K$, where $K$ is defined in \eqref{eq:K}.
\end{Asm}
Roughly speaking, Assumption \ref{Asm:2} requires the number of active abundances at each HS pixel to be no greater than the degrees of distinctiveness of $\bar{\bA}'$.
It should be noted that
\begin{equation} \label{eq:imply_sparse_s}
%\| \bar{\bs}_i' \|_0 \leq K \quad \Longrightarrow \quad 
%\| \bar{\bs}_j \|_0 \leq K \text{~for all $j \in \setL_i$;}
\supp(\bar{\bs}_j) \subseteq \supp(\bar{\bs}_i'), \text{~for all $j \in \setL_i$.}
\end{equation}
Here, the notation $\supp$ is defined as $\supp(\bx) = \{ i \mid x_i \neq 0 \}$.
Eq.~\eqref{eq:imply_sparse_s} can be easily verified from $\bar{\bs}_i'= \bar{\bS}_{\setL_i} \bg_i$, with $\bg_i > 0$ and $\bar{\bS}_{\setL_i} \geq \bzero$.
There is an interesting consequence with Assumption \ref{Asm:2}.
Without Assumption \ref{Asm:2}, the linear system solution to \eqref{eq:bYm_rewrite} with respect to $\bS$ is non-unique when $\bar{\bA}'$ is fat.
With Assumption \ref{Asm:2}, we can  express as \eqref{eq:bYm_rewrite}
\begin{equation} \label{eq:bymi_temp}
\by_{{\rm M},j} = \bar{\bA}_{\setJ_i}' \bar{\bs}_j^{\setJ_i}, \quad j \in \setL_i,
\end{equation}
and for all $i=1,\ldots,\Lh$,
where $\setJ_i = \supp(\bar{\bs}_i')$.
Since $|\setJ_i| = \| \bar{\bs}_i' \|_0 \leq K$, the matrix $\bar{\bA}_{\setJ_i}'$ has full column rank.
Hence, the linear system solution to \eqref{eq:bymi_temp} with respect to ${\bs}_j^{\setJ_i}$ is unique---assuming that we know the sparsity pattern indicated by $\setJ_i$.
The above observation, first made in \cite{hsr_recovery_ssp2018}, will be key in our CoSMF recovery analysis.

In addition to sparsity, we also adopt the widely-used pure-pixel assumption in HS unmixing \cite{ma2013signal} (also known as the separability assumption in machine learning \cite{Gillis14}).
\begin{Asm} \label{Asm:3}
	There exists an index set $\setK$ such that the spatially-decimated true abundance matrix $\bar{\bS}'$ satisfies $\bar{\bS}'_{\setK} = \bI$.
\end{Asm}
The above pure-pixel assumption means that there exist HS pixels that contain purely one material; this can be seen from $[ \bY_{{\rm H}} ]_\setK = \bar{\bA} \bar{\bS}'_{\setK} = \bar{\bA}$.
%Note that whether the pure-pixel assumption holds, or can serve as a reasonable approximation, is scenario-dependent.
The pure-pixel assumption is reasonable if the HS spatial resolution relative to the scene is not too coarse such that pure (or near-pure) pixels of each material appear at least once in the scene.

We need one more assumption to pin down the recovery condition of CoSMF.
Define 
\begin{align*}
\epsilon_{ji} & = \min_{1 \leq k \leq M} \bigg\{ \frac{1-\bar{a}_{kj}}{1-N \bar{a}_{ki}} ~ \bigg| ~ \bar{a}_{ki} < \frac{1}{N} \bigg\}, \\
\epsilon & = \max_{1 \leq j, i \leq N, i \neq j} \epsilon_{ji}.
\end{align*}
Note that the above definition implicitly assumes $\bar{a}_{ki} < 1/N$ for some $k$, and that $\epsilon \geq 0$.
\begin{Asm} \label{Asm:4}
	It holds that $\min_{1 \leq k \leq M} \bar{a}_{ki} < 1/N$ for all $i$, and that $\epsilon < 1/(4N)$.
\end{Asm}
A simple way to interpret Assumption \ref{Asm:4} is as follows.
We can find an HS spectral band, indexed by $k$, such that the spectral component of one material, $\bar{a}_{kj}$, is much stronger than that of another, $\bar{a}_{ki}$.
This dominant condition is more significant if $\epsilon$ is smaller.
Assumption \ref{Asm:4} may look strong, but it is necessary for CoSMF to guarantee good provable recovery results; this will be discussed later.
Moreover, we can justify by considering the following result.
\begin{Lemma} \label{lem:prob_A}
	If the elements $\bar{a}_{ij}$ of the true endmember matrix $\bar{\bA}$ are  independent and follow the $[0,1]$-uniform distribution, Assumption \ref{Asm:4} holds with probability at least $1- N (N-1)\exp(-\frac{M}{8N^2})$.
\end{Lemma}
\ifconfver
	The proof of Lemma~\ref{lem:prob_A} is shown in the extended version of this paper \cite{temp}.
\else
	The proof of Lemma~\ref{lem:prob_A} is shown in Appendix~\ref{app:lem:prob_A}.
\fi

We now present the main result.
\begin{Theorem} \label{thm:main}
	Suppose Assumptions 1--4 hold. 
	Also, suppose $N \geq 2$.
	Then, any solution $(\bA,\bS)$ to the CoSMF problem \eqref{eq:CoSMF} satisfies
	\[
	\| \bar{\bA} \bar{\bs}_j - \bA \bs_j \|_2 \leq \epsilon \cdot \sigmax(\bar{\bA}) \sqrt{1+\kappa^2} \left(4 + \frac{2}{\gamma_j} \right) C,
	\]
	 where 
	\begin{align}
	\kappa & = \max_{\substack{\setJ \subseteq \{1,\ldots,N\}, \\ 1 \leq |\setJ| \leq N}}  \frac{\sigmax\left(\bar{\bA}'_{\{1,\ldots,N\} \setminus \setJ} \right)}{\sigmin(\bar{\bA}'_\setJ)},
	\label{eq:kappa_def} \\
	C & = \left\{ \begin{array}{ll} N/2, & K \geq N/2 \\ \sqrt{K(N-K)}, & K < N/2 \end{array} \right. \label{eq:C_def} \\
	\gamma_j & = \max\{ g_{ji} \mid i=1,\ldots,\Lh,  \setL_i \ni j  \}.
	\label{eq:gamma_def}
	\end{align}
\end{Theorem}
We will prove Theorem \ref{thm:main} in the next section.
Theorem \ref{thm:main} is not an exact recovery result.
In fact, we will show in Section~\ref{sec:counter_example} that a counter example for exact recovery exists.
Theorem \ref{thm:main} shows, for the first time, that CoSMF is theoretically guaranteed to yield certain recovery accuracies.
It also suggests how the recovery error may scale with the problem parameters.
The most notable one is $\epsilon$.
If the spectral dominant condition with the true endmembers in Assumption~\ref{Asm:4} is strong such that $\epsilon$ is small, the recovery error will be small.
Another parameter is $\kappa$, which looks similar to the condition number (i.e.,  $\sigmax({\bA})/\sigmin(\bA)$) in the study of linear system sensitivity.
We expect $\kappa$ to be small if the spectrally-decimated endmember matrix $\bar{\bA}'$ exhibits good distinctiveness.
%Conversely it also implies that the spectrally-decimated endmembers look similar, 
%As can be seen in (XX), the recovery error scales with the parameter $\kappa$, which is somehow reminiscent of the condition number of linear systems.
%Intuitively, if the endmembers show significant differences, $\kappa$ is expected to become small.

An interesting aspect we should highlight is that the CoSMF problem \eqref{eq:CoSMF} does not exploit the sparsity and pure-pixel problem structures.
It is a plain matrix factorization, utilizing only the $[0,1]$ and unit-simplex properties of the endmembers and abundances, respectively.
Yet, the sparsity and pure-pixel conditions play a hidden role in endowing CoSMF with some recovery characteristics.
Those characteristics will be revealed in the proof of Theorem \ref{thm:main} in the next section.

It should also be mentioned that Theorem \ref{thm:main} shows a recovery error bound that is applicable to any globally optimal solution to the CoSMF problem \eqref{eq:CoSMF}.
In practice, if we initialize the algorithm well such that the algorithm stands a good chance to converge to a solution close to the ground-truth $(\bar{\bA},\bar{\bS})$ (or $(\bar{\bA} \bm\Pi, \bm\Pi^T\bar{\bS})$ for any permutation matrix $\bm\Pi$), 
we should expect that the recovery error be smaller than the theory predicts.

%\section{Proof of Theorem \ref{thm:main}}
\section{Proof of Theorem 1}

Since we assume no noise, the CoSMF problem \eqref{eq:CoSMF} is the same as
\begin{subequations} \label{eq:prob_main_feas}
\begin{align}
{\rm find} & ~ \bA \in \setA, \bS \in \setS \\
{\rm s.t.} & ~ \bYh = \bA \bS', \label{eq:prob_main_feas_bYh}  \\ 
& ~ \bYm = \bA' \bS,
\end{align}
\end{subequations}
where $\bA' = \bF \bA, \bS' = \bS \bG$.
Let us simply denote $(\bA,\bS)$ as an arbitrary solution to problem \eqref{eq:prob_main_feas}.
From \eqref{eq:bYh_rewrite} and \eqref{eq:prob_main_feas_bYh} we have
\[
\bar{\bA}\bar{\bS}' = \bA \bS'.
\]
Since we assume that $\bar{\bA}$ and $\bar{\bS}'$ have full column and row rank, respectively (Assumption \ref{Asm:1}), by basic matrix analysis results we know that $(\bA,\bS')$ must satisfy
\begin{equation} \label{eq:there_R}
\bA = \bar{\bA} \bR^{-1}, \quad \bS' = \bR \bar{\bS},
\end{equation}
for some nonsingular $\bR$.
This leads to
\begin{align}
\| \bar{\bA} \bar{\bs}_j - \bA\bs_j \|_2 & = \| \bar{\bA} \bar{\bs}_j - \bar{\bA} \bR^{-1} \bs_j \|_2 \nonumber \\
& \leq \sigmax(\bar{\bA}) \| \bar{\bs}_j - \bR^{-1} \bs_j \|_2
\label{eq:bnd_AS}
\end{align}

The next problem is to show a bound on $\| \bar{\bs}_j - \bR^{-1} \bs_j \|_2$.
\begin{Prop} \label{prop:bnd_S}
	Let $\bm \Pi \in \Rbb^{N \times N}$ be a permutation matrix.
	Let $\tilde{\bR} = \bm\Pi \bR$, and let $\rho = \max_{i \neq j } |\tilde{r}_{ij} |$.
	Suppose $N \geq 2$ and
	\begin{equation} \label{eq:beta_def}
	\beta := \min_{\substack{\setI \subseteq \{1,\ldots,N\}, \\ N-K \leq |\setI| \leq N-1}} 
	\sigmin(\tilde{\bR}_\setI^\setI) > 0.
	\end{equation}
	Under Assumption \ref{Asm:2}, it holds that 
	\begin{equation} \label{eq:bnd_S}
	\| \bar{\bs}_j - \bR^{-1} \bs_j \|_2 \leq 
		\sqrt{1+ \kappa^2}  \frac{\rho C}{\beta} 
		\left( \frac{1}{\sigmin(\tilde{\bR})} + \frac{1}{\gamma_j}   \right),
	\end{equation}
	where $\kappa$, $C$ and $\gamma_j$ are  defined in \eqref{eq:kappa_def}, \eqref{eq:C_def}, and \eqref{eq:gamma_def}, respectively.
	
\end{Prop}
\ifconfver
The proof of Proposition \ref{prop:bnd_S} is shown in the extended version of this paper \cite{temp}.
\else
The proof of Proposition \ref{prop:bnd_S} is shown in Appendix \ref{app:prop:bnd_S}.
\fi
%The proof of Proposition \ref{prop:bnd_S} is shown in the extended version of this paper \cite{temp}.
Proposition \ref{prop:bnd_S} reveals that if $\tilde{\bR}$, a row-wise permutation of $\bR$, is close to a diagonal matrix, the error $\| \bar{\bs}_j - \bR^{-1} \bs_j \|_2$ will be small.

We therefore study the structures of $\bR$, seeing if a near-diagonal $\tilde{\bR}$ can be found.
Here is the first result.
\begin{Fact} \label{fact:pure_pix}
	Under Assumption \ref{Asm:3} we have $\bR \in \setU^{N \times N}$, or equivalently,
	\[
	0 \leq r_{ij} \leq 1, \quad r_{ii} = 1 - \textstyle \sum_{k \neq i} r_{ki},
	\]
	for all $i,j$.
\end{Fact}
The proof of Fact \ref{fact:pure_pix} is simple:
From \eqref{eq:there_R} we have ${\bS}'_{\setK} = \bR \bar{\bS}_\setK = \bR$.
Since $\bS' \in \setU^{N \times L}$ (which follows the same argument as $\bar{\bS} \in  \setU^{N \times L}$ in the preceding section), we obtain $\bR \in \setU^{N \times N}$.
With Fact \ref{fact:pure_pix}, we further show the following result.
\begin{Prop} \label{prop:bnd_A}
	Under Assumption \ref{Asm:4} and the result in Fact \ref{fact:pure_pix}, there exists a permutation matrix $\bm \Pi$ such that $\tilde{\bR} = \bm \Pi \bR$ satisfies $\tilde{r}_{ij} \leq \epsilon$ for all $i \neq j$. 
	Here, the requirement with $\epsilon$ in  Assumption \ref{Asm:4} can be relaxed as $\epsilon < 1/N$.
\end{Prop}
\ifconfver
	The proof of Proposition \ref{prop:bnd_A} is shown in the extended version of this paper \cite{temp}.
\else
	The proof of Proposition \ref{prop:bnd_A} is shown in Appendix \ref{app:prop:bnd_A}.
\fi
By applying Propositions \ref{prop:bnd_S} and \ref{prop:bnd_A} to \eqref{eq:bnd_AS}, 
%with $\rho \leq \epsilon$, 
the proof of Theorem \ref{thm:main} is almost complete.
The remaining problem is to bound $\sigmin(\tilde{\bR})$ and $\beta$ in \eqref{eq:bnd_S}. 
We use the following result which is the consequence of a singular value bound for strictly diagonally dominant matrices \cite{varah1975lower}.
\begin{Fact} \label{fact:sddm}
	Suppose $\tilde{\bR} = \bm\Pi \bR$, where $\bm\Pi$ is a permutation matrix and $\bR$ lies in $\setU^{N \times N}$, satisfies $\tilde{r}_{ij} \leq 1/(4N)$ for all $i \neq j$.
	Then, it is true that $\sigmin(\tilde{\bR}_\setI^\setI) \geq 1/2$ for any non-empty $\setI \subseteq \{1,\ldots, N\}$.
\end{Fact}
\ifconfver
	We show Fact \ref{fact:sddm} in the extended version of this paper \cite{temp}.
\else
	We show Fact \ref{fact:sddm} in Appendix \ref{app:fact:sddm}. 
\fi
Applying the above result completes the proof of Theorem \ref{thm:main}.

\section{Is Exact Recovery Possible?}
\label{sec:counter_example}

One may wonder if CoSMF can guarantee exact recovery under our assumptions.
We argue that this is impossible in general.
%, and we base our argument on construction of a counter example.
Consider the following counter example:
 $M= 3, \Mm= 1, N= 3, L= 6$, 
\[
\bar{\bA} =
\begin{bmatrix}
1-\rho & \rho & 0 \\
\rho & 1-\rho & 0 \\
0   & 0   & 1
\end{bmatrix}, \quad
\bar{\bS} =
\begin{bmatrix}
1 & 1 & 0 & 0 & 0 & 0 \\
0 & 0 & 1 & 1 & 0 & 0 \\
0 & 0 & 0 & 0 & 1 & 1
\end{bmatrix},
\]
$\bF = [~1, 1, 1 ~]$,
$\setL_1 = \{ 1, 2\}, \setL_2 = \{ 3, 4\}, \setL_3 = \{ 5, 6 \}$, $\bg_1 = \bg_2 = \bg_3 = [ 0.5, 0.5 ]^T$,
$0 \leq \rho < 0.5$.
This instance satisfies Assumptions~\ref{Asm:1}--\ref{Asm:3}, with $K= 1$.
It can be verified that $\bA = \bI$,
\[ 
\bS = 
\begin{bmatrix}
1-\rho +\alpha_1 & 1-\rho - \alpha_1 & \rho +\alpha_2 & \rho - \alpha_2 & 0 & 0 \\
\rho-\alpha_1 & \rho + \alpha_1 & 1-\rho-\alpha_2 & 1-\rho+\alpha_2 & 0 & 0 \\
0 & 0 & 0 & 0 & 1 & 1
\end{bmatrix}
\]
is a feasible solution to problem \eqref{eq:prob_main_feas} for any $-\rho \leq \alpha_i \leq \rho$, $i=1,2$;
we omit the verification details owing to space limitation.
We see that 
\begin{equation} \label{eq:counter_exa_bnd}
\| \bar{\bA} \bar{\bs}_1 - \bA \bs_1 \|_2 = \sqrt{2} | \alpha_1 | \leq \sqrt{2} \rho.
\end{equation}
Also, the above error bound is achievable (choose $\alpha_1 = \rho$).
This demonstrates that exact recovery is impossible except for the special case of $\rho= 0$.
This counter example also helps explain why Assumption \ref{Asm:4} is necessary.
The error bound \eqref{eq:counter_exa_bnd} can only be reduced by decreasing $\rho$.
On the other hand, the requirement of small $\epsilon$ in  Assumption \ref{Asm:4}  is  the same as forcing $\rho$ to be small.

\section{Conclusion}

We proved the sufficient recovery guarantees of a CoSMF problem for HSR.
Our analysis revealed that the abundance sparsity, the existence of pure pixels, and some spectral endmember dominant property provide the sufficient conditions for CoSMF to yield good recovery guarantees.
% and the pure-pixel condition are important for pinning down the sufficient recovery guarantees of CoSMF.
%play a hidden role in allowing the CoSMF to yield some good recovery guarantees.

\ifconfver

\else

\section*{Appendix}
\renewcommand{\thesubsection}{\Alph{subsection}}

\subsection{Proof of Lemma \ref{lem:prob_A}}
\label{app:lem:prob_A}

Let $\mathcal E$ denote the event that Assumption \ref{Asm:4} is violated.
The occurrence of $\mathcal E$ means that there exists an index pair $(i,j)$, with $1\leq i,j\leq N$ and $i\neq j$, such that either $\min_{1 \leq k \leq M} \bar{a}_{ki} \geq 1/N$  or $\epsilon_{ji}\geq 1/(4N)$ holds.
We can represent $\mathcal E$ by
\[
\mathcal E = \bigcup_{1 \leq j, i \leq N, j \neq i } \mathcal E_{ji},
\]
where 
\begin{align*}
\mathcal E_{ji} 
& =  \left\{  \min_{1 \leq k \leq M} \bar{a}_{ki} \geq 1/N \text{~or~} \epsilon_{ji}\geq 1/(4N)  \right\} \\
& = \bigcap_{1 \leq k \leq M } \left\{  \bar{a}_{ki} \geq \frac{1}{N} \text{~or~} \frac{1 - \bar{a}_{kj}}{1- N \bar{a}_{ki} } \geq \frac{1}{4N}  \right\}.
\end{align*}
It can be verified that
\[
\left\{  \bar{a}_{ki} \geq \frac{1}{N} \text{~or~} \frac{1 - \bar{a}_{kj}}{1- N \bar{a}_{ki} } \geq \frac{1}{4N}  \right\} =
\left\{
 1- \bar{a}_{kj} \geq \frac{1  -  N \bar{a}_{ki} }{4N}
\right\}.
\]
Recall that $\bar{a}_{kj}$ and $\bar{a}_{ki}$ are independent and $[0,1]$-uniformly distributed random variables. We can easily show that
\[
{\rm P}\left( \left\{
1- \bar{a}_{kj} \geq \frac{1  -  N \bar{a}_{ki} }{4N}
\right\}   \right) = 1 - \frac{1}{8N^2}.
\]
It follows that 
\begin{align*}
{\rm P}(\mathcal E_{ji}) & = \prod_{1 \leq k \leq M} {\rm P}\left( \left\{
1- \bar{a}_{kj} \geq \frac{1  -  N \bar{a}_{ki} }{4N}
\right\}   \right)  \\
& = \left( 1 - \frac{1}{8N^2} \right)^M \\
& \leq \exp\left(-\frac{ M}{8N^2} \right),
\end{align*}
where the first equation uses the independence of the random variables $\bar{a}_{ki}$'s with respect to $k$;
the last equation holds because  $1-x\leq \exp(-x)$.
Furthermore, by applying the above inequality to ${\rm P}(\mathcal E) \leq \sum_{j \neq i} {\rm P}(\mathcal E_{ji})$, we obtain ${\rm P}(\mathcal E) \leq N(N-1) \exp(-M/(8N^2))$.
This completes the proof.

\subsection{Proof of Proposition \ref{prop:bnd_S}}
\label{app:prop:bnd_S}

From the preceding development we have the following equalities:
\begin{align*}
\bA' & = \bar{\bA}' \tilde{\bR}^{-1} \bm\Pi, \quad 
\bar{\bA}' \bar{\bS}_{\setL_i}  = \bA' \bS_{\setL_i}, \quad \bar{\bs}_i' = \tilde{\bR}^{-1} \bm\Pi \bs_i',
\quad
\bar{\bs}_i' = \bar{\bS}_{\setL_i} \bg_i, \quad
{\bs}_i' = {\bS}_{\setL_i} \bg_i.
\end{align*}
To keep the notations simple, within this proof we re-define $\bar{\bS} = \bar{\bS}_{\setL_i}$, ${\bS} = \bm\Pi {\bS}_{\setL_i}$, $\bar{\bs}'= \bar{\bs}_i'$, ${\bs}'= \bm\Pi {\bs}_i'$, $\bg_i = \bg$, $L= |\setL_i|$, 
%$\bA = \bA \bm\Pi$, 
$\bA' = \bA' \bm\Pi^T$.
Doing so simplifies the above equations to
\begin{align}
\bA' & = \bar{\bA}' \tilde{\bR}^{-1}, \label{eq:bnd_S_t1o} \\
\bar{\bA}' \bar{\bS} & = \bA' \bS, \label{eq:bnd_S_t1a} \\
\bar{\bs}' & = \tilde{\bR}^{-1} \bs', \label{eq:bnd_S_t1b} \\
\bar{\bs}' & = \bar{\bS} \bg, \quad {\bs}'  = {\bS} \bg,   \label{eq:bnd_S_t1c}
\end{align}
and our aim becomes proving a bound on $\| \bar{\bs}_j - \tilde{\bR}^{-1} \bs_j \|_2$.
Let $\setJ = \supp(\bar{\bs}')$ and $\setI = \{1,\ldots,N\} \setminus \setJ$.
%Note that $|\setJ| \leq K$, and 
As $\bar{\bs}_j^\setI = \bzero$ for all $j$ (cf. \eqref{eq:imply_sparse_s}),
Eq.~\eqref{eq:bnd_S_t1a} can be re-expressed as
\begin{align} \label{eq:bnd_S_t2}
\bar{\bA}'_\setJ \bar{\bs}_j^\setJ & = \bA' \bs_j, \quad j=1,\ldots,L.
\end{align}
Also, by Assumption \ref{Asm:2} we have $|\setJ| \leq K = {\rm krank}(\bar{\bA}')$.
This implies that $\bar{\bA}'_\setJ$ has full column rank and therefore has $\sigmin(\bar{\bA}'_\setJ) > 0$.

Firstly, we show that
\begin{equation} \label{eq:bnd_S_key1}
\| \bar{\bs}_j - \tilde{\bR}^{-1} \bs_j \|_2  \leq \sqrt{1+ \kappa^2_\setJ}
\| [ \tilde{\bR}^{-1} \bs_j ]^\setI \|_2,
\end{equation}
where $\kappa_\setJ = \sigmax(\bar{\bA}'_\setI)/\sigmin(\bar{\bA}'_\setJ)$.
The proof is as follows.
Since, by \eqref{eq:bnd_S_t1o},
\[
\bA' \bs_j = \bar{\bA}' \tilde{\bR}^{-1} \bs_j  = \bA'_\setJ [ \tilde{\bR}^{-1} \bs_j ]^\setJ + \bA'_\setI [ \tilde{\bR}^{-1} \bs_j ]^\setI,
\]
Eq.~\eqref{eq:bnd_S_t2} can be re-organized as
\begin{align*} 
\bar{\bA}'_\setJ \left( \bar{\bs}_j^\setJ - [ \tilde{\bR}^{-1} \bs_j ]^\setJ \right) & =  \bA'_\setI [ \tilde{\bR}^{-1} \bs_j ]^\setI.
\end{align*}
Applying the Euclidean norm on both sides of the above equation, we obtain 
\begin{equation} \label{eq:bnd_S_t3}
\sigmin(\bar{\bA}'_\setJ) \| \bar{\bs}_j^\setJ - [ \tilde{\bR}^{-1} \bs_j ]^\setJ \|_2 \leq 
\sigmax(\bA'_\setI) \| [ \tilde{\bR}^{-1} \bs_j ]^\setI \|_2.
\end{equation}
It follows that
\begin{align}
\| \bar{\bs}_j - \tilde{\bR}^{-1} \bs_j \|_2^2 
	& = \| [ \bar{\bs}_j - \tilde{\bR}^{-1} \bs_j ]^\setI \|_2^2 + \| [ \bar{\bs}_j - \tilde{\bR}^{-1} \bs_j ]^\setJ \|_2^2 \nonumber \\
	& \leq \| [ \tilde{\bR}^{-1} \bs_j ]^\setI \|_2^2 + \kappa_\setJ^2 \| [ \tilde{\bR}^{-1} \bs_j ]^\setI \|_2^2, \nonumber
\end{align}
which is due to $\bar{\bs}_j^\setI = \bzero$ and \eqref{eq:bnd_S_t3}.
This completes the proof of \eqref{eq:bnd_S_key1}.

Secondly, we show that
\begin{equation} \label{eq:bnd_S_key2}
\| [ \tilde{\bR}^{-1} \bs_j ]^\setI \|_2 \leq \frac{1}{\sigmin(\tilde{\bR}_\setI^\setI)} 
\left( \frac{ \| \tilde{\bR}_\setJ^\setI \|_2 }{\sigmin(\tilde{\bR})}  + \| \bs_j^\setI \|_2 \right).
\end{equation}
To prove it, observe
\[
\bs_j^\setI = [  \tilde{\bR} \tilde{\bR}^{-1} \bs_j ]^\setI 
= \bR_\setI^\setI [ \tilde{\bR}^{-1} \bs_j ]^\setI + \bR_\setJ^\setI [ \tilde{\bR}^{-1} \bs_j ]^\setJ.
\]
By re-arranging the above equation as
\begin{equation} \label{eq:bnd_S_t4}
\tilde{\bR}_\setI^\setI [ \tilde{\bR}^{-1} \bs_j ]^\setI 
 = \bs_j^\setI - \tilde{\bR}_\setJ^\setI [ \tilde{\bR}^{-1} \bs_j ]^\setJ,
\end{equation}
and by taking the Euclidean norm on both sides of \eqref{eq:bnd_S_t4}, we have
\begin{align}
\| \tilde{\bR}_\setI^\setI [ \tilde{\bR}^{-1} \bs_j ]^\setI  \|_2 & \geq
	\sigmin(\tilde{\bR}_\setI^\setI) \| [ \tilde{\bR}^{-1} \bs_j ]^\setI  \|_2, 
%	\nonumber \\
%	& \geq \sigmin(\tilde{\bR}) \| [ \tilde{\bR}^{-1} \bs_j ]^\setI  \|_2, 
	\label{eq:bnd_S_t4a}  \\
\| \bs_j^\setI - \tilde{\bR}_\setJ^\setI [ \tilde{\bR}^{-1} \bs_j ]^\setJ \|_2 & 
	\leq  \| \bs_j^\setI \|_2 + \| \tilde{\bR}_\setJ^\setI \|_2 \| [ \tilde{\bR}^{-1} \bs_j ]^\setJ \|_2      \nonumber \\
	& \leq \| \bs_j^\setI \|_2 + \| \tilde{\bR}_\setJ^\setI \|_2 \| \tilde{\bR}^{-1} \bs_j  \|_2    \nonumber  \\
	& \leq \| \bs_j^\setI \|_2 + \| \tilde{\bR}_\setJ^\setI \|_2 \| \tilde{\bR}^{-1} \|_2 \| \bs_j  \|_2  \nonumber \\
	& \leq \| \bs_j^\setI \|_2 + \frac{\| \tilde{\bR}_\setJ^\setI \|_2}{\sigmin(\tilde{\bR})},
	\label{eq:bnd_S_t4b}
\end{align}
respectively.
%Note that \eqref{eq:bnd_S_t4a} is due to $\sigmin(\tilde{\bR}_\setI^\setI) \geq \sigmin(\tilde{\bR})$, which is a direct consequence of the singular value result $\sigma(\bB) = \min_{\| \bx \|_2 = 1} \| \bB \bx \|_2$ for full column-rank $\bB$.
Note that, in \eqref{eq:bnd_S_t4b}, we have applied $\| \bs_j \|_2  \leq \| \bs_j \|_1 = 1$ for $\bs_j \in \setU^N$.
By applying \eqref{eq:bnd_S_t4a}--\eqref{eq:bnd_S_t4b} to \eqref{eq:bnd_S_t4}, we obtain \eqref{eq:bnd_S_key2}.

Thirdly, we show that
\begin{align}
\| \bs_j^\setI \|_2 & \leq \frac{\rho}{g_j} \sqrt{N-|\setJ|},   \label{eq:bnd_S_key3a} \\
\| \tilde{\bR}_\setJ^\setI \|_2 & \leq \rho \sqrt{|\setJ|(N- |\setJ|)}. \label{eq:bnd_S_key3b}
\end{align}
Eq.~\eqref{eq:bnd_S_key3a} is the consequence of 
\begin{equation} \label{eq:bnd_S_t7}
s_{ij} \leq \rho/g_j,
\end{equation}
which is shown as follows.
From \eqref{eq:bnd_S_t1b} we see that
\begin{align*}
s_i' & = [ \tilde{\bR} \bar{\bs}' ]^i = \tilde{r}_{ii} \bar{s}_i' + \sum_{k \neq i} \tilde{r}_{ik} \bar{s}_k'
\end{align*}
for any $i$. For $i \in \setI$, the above equation yields
\begin{align} \label{eq:bnd_S_t5}
s_i' & \leq \max_{j \neq i} \tilde{r}_{ij} \sum_{k \neq i} \bar{s}_k' \leq  \rho,
\end{align}
which is due to $\bar{s}_i' = 0$, $\bar{\bs}' \geq \bzero$, and $\sum_{k} \bar{s}_k' = 1$.
On the other hand, we see from \eqref{eq:bnd_S_t1c} that
\begin{align} \label{eq:bnd_S_t6}
s_i' & = \sum_{k=1}^L g_k s_{ik} \geq g_j s_{ij}, 
\end{align}
for any $i,j$.
Combining \eqref{eq:bnd_S_t5} and \eqref{eq:bnd_S_t6} leads to \eqref{eq:bnd_S_t7}.
Eq.~\eqref{eq:bnd_S_key3b} is simply due to 
%recall $|\tilde{r}_{ij}| \leq \rho$ for any $i \neq j$.
%We then have
\[
\| \tilde{\bR}_\setJ^\setI \|_2^2 \leq \| \tilde{\bR}_\setJ^\setI \|_F^2 = \sum_{i \in \setI} \sum_{j \in \setJ} |\tilde{r}_{ij} |^2 \leq \rho^2 |\setI||\setJ| = \rho^2 (N- |\setJ|)|\setJ|.
\]

Finally, we combine all of the above results. 
From \eqref{eq:bnd_S_key1}, \eqref{eq:bnd_S_key2}, \eqref{eq:bnd_S_key3a} and \eqref{eq:bnd_S_key3b}, we have
\begin{align*} 
\| \bar{\bs}_j - \tilde{\bR}^{-1} \bs_j \|_2  & \leq \sqrt{1+ \kappa^2_\setJ}
\frac{\rho}{\sigmin(\tilde{\bR}_\setI^\setI)} 
\left( \frac{ \sqrt{|\setJ|(N- |\setJ|)}  }{\sigmin(\tilde{\bR})}  + \frac{\sqrt{N- |\setJ|}}{g_j} \right) \\
& \leq \sqrt{1+ \kappa^2_\setJ}
\frac{\rho}{\beta} 
\left( \frac{1}{\sigmin(\tilde{\bR})}  + \frac{1}{g_j} \right) \sqrt{|\setJ|(N- |\setJ|)},
\end{align*}
where $\beta$ is given by \eqref{eq:beta_def}.
The final outcome in \eqref{eq:bnd_S} is a minor furnishing of the above inequality.
Specifically, we further bound
\[
\kappa_\setJ \leq \kappa = \max_{\substack{\setJ \subseteq \{1,\ldots,N\}, \\ 1 \leq |\setJ| \leq N}}  \frac{\sigmax(\bA'_\setI)}{\sigmin(\bA'_\setJ)} 
\]
and use the fact that $\kappa$ remains the same if we permute the columns of $\bA'$ (recall that in the beginning we re-define $\bA'$ as a column-permuted version of the original $\bA'$).
In addition, it can be verified that, for $N \geq 2$,
\[
\max_{1 \leq |\setJ| \leq K} \sqrt{|\setJ|(N- |\setJ|)} \leq C = \left\{ \begin{array}{ll} N/2, & K \geq N/2 \\ \sqrt{K(N-K)}, & K < N/2 \end{array} \right.
\]
Furthermore, when we use back the notations in the main content, we get
\begin{align*} 
\| \bar{\bs}_j - {\bR}^{-1} \bs_j \|_2  
& \leq \sqrt{1+ \kappa^2}
\frac{\rho}{\beta} 
\left( \frac{1}{\sigmin(\tilde{\bR})}  + \frac{1}{g_{ji}} \right) C,
\end{align*}
for all $j$ and for any $i$ such that $j \in \setL_i$.
By choosing $i$ such that $1/g_{ji}$ is the smallest, we arrive at the final expression  \eqref{eq:bnd_S}. 
The proof is complete.

\subsection{Proof of Proposition \ref{prop:bnd_A}}
\label{app:prop:bnd_A}

Firstly, we argue that $\tilde{\bR}$ can be made to satisfy
\begin{equation} \label{eq:p_ii}
\tilde{r}_{ii} = \max_{j \geq i} \tilde{r}_{ij}.
\end{equation}
Also, such a $\tilde{\bR}$ satisfies
\begin{equation} \label{eq:p_ij}
\tilde{r}_{ki} \leq \epsilon, ~ 1\leq k \leq i-1
\quad \Longrightarrow \quad \tilde{r}_{ii} \geq 1/N.
\end{equation}
%then $p_{ii} \geq 1/N$.
By the same spirit as partial pivoting in linear systems, we can re-order the rows of $\bR$ such that the re-ordered $\bR$, which is $\tilde{\bR}$, satisfies \eqref{eq:p_ii}.
To prove \eqref{eq:p_ij}, note that $\tilde{\bR} \in \setU^{N \times N}$ (this is easy to verify).
By the implication $\bx \in \setU^n \Longrightarrow \max_{1 \leq j \leq n} x_j \geq 1/N$ and  the assumption $ \epsilon < 1/N$, we observe that \eqref{eq:p_ij} must be true.

Secondly, we use induction to show that $\tilde{r}_{ij} \leq \epsilon$ for all $i \neq j$.
Let $i$ be fixed. Suppose, for $i \geq 2$, it holds that
\begin{equation} \label{eq:p_ij_2}
\tilde{r}_{lj} \leq \epsilon,  \quad {\rm for}~1 \leq l \leq i-1, 1\leq j \leq N, j \neq l. 
\end{equation}
%Note that \eqref{eq:p_ij_2} is trivially true for $i=1$.
We want to show that 
%$\tilde{r}_{ij} \leq \epsilon$ for all $j \neq i$.
\[
\tilde{r}_{ii} \geq 1/N, \quad \tilde{r}_{ij} \leq \epsilon, \quad {\rm for}~1 \leq j \leq N, j \neq i.
\]
The inequality $\tilde{r}_{ii} \geq 1/N$ is obtained by \eqref{eq:p_ij}.
The proof of $\tilde{r}_{ij}$ for $j \neq i$ goes as follows.
Fixing $j \neq i$, Assumption \ref{Asm:4} implies that there exists an index $k$ such that
\begin{align}
\bar{a}_{ki} & < 1/N, \label{eq:prop:bnd_A_t1} \\
\epsilon N \bar{a}_{ki} + 1 - \epsilon & \leq \bar{a}_{ki}. \label{eq:prop:bnd_A_t2}
\end{align}
Consider \eqref{eq:there_R}
\begin{equation} \label{eq:AAP}
\bar{\bA} = \bA\bPi \tilde{\bR},
\end{equation}
and assume $\bPi=\bI$ without loss of generality
From the $(k,i)$th element of \eqref{eq:AAP}, we get
\begin{equation} \label{eq:aki}
\bar{a}_{ki} = \tilde{r}_{1,i} {a}_{k,1} + \cdots + \tilde{r}_{N,i} {a}_{k,N}
\geq \tilde{r}_{ii} {a}_{k,i} \geq \frac{1}{N} {a}_{ki},
\end{equation}
which is due to $\bA \geq \bzero$, $\tilde{r}_{li} \geq 0$, and $\tilde{r}_{ii} \geq 1/N$.
Eq.~\eqref{eq:aki} and \eqref{eq:prop:bnd_A_t1} imply 
\[ a_{ki} < 1. \]
Moreover, from the $(k,j)$th element of \eqref{eq:AAP} we derive
\begin{align}
\bar{a}_{kj} & = \tilde{r}_{ij}  {a}_{ki}  + \textstyle \sum_{l \neq i} \tilde{r}_{lj} {a}_{kl}  
%\leq  \tilde{r}_{ij}  {a}_{ki} + \sum_{l \neq i} \tilde{r}_{lj} 
%\nonumber \\
%& 
\leq \tilde{r}_{ij}  {a}_{ki} + (1- \tilde{r}_{ij}),
\label{eq:akj}
\end{align}
where is due to $\bA \leq \bone$, $\tilde{r}_{lj} \geq 0$, and $\sum_{l} \tilde{r}_{lj} = 1$.
By putting \eqref{eq:aki} and \eqref{eq:akj} into the left-hand side and right-hand side of \eqref{eq:prop:bnd_A_t2}, respectively, we have
\[
\epsilon a_{ki} + (1-\epsilon) \leq \tilde{r}_{ij}  a_{ki} + (1-\tilde{r}_{ij}).
\] 
By re-arranging the terms in the above inequality, we further write
\[
\epsilon (a_{ki} - 1) \leq \tilde{r}_{ij} (a_{ki} - 1).
\]
Since $0 \leq a_{ki} < 1$, the above inequality is equivalent to $\tilde{r}_{ij} \leq \epsilon$.
Hence we have shown that $\tilde{r}_{ij} \leq \epsilon$ for all $j \neq i$.
By increasing $i$ by one, and repeating the above step, we conclude that $\tilde{r}_{ij} \leq \epsilon$ for all $i \neq j$.
The proof is complete.

\subsection{Proof of Fact \ref{fact:sddm}}
\label{app:fact:sddm}

Given a matrix $\bB \in \Rbb^{n \times n}$, define
\begin{align*}
c_i(\bB) & = |b_{ii}| - \sum_{k \neq i} | b_{ki} |, \\
d_i(\bB) & = |b_{ii}| - \sum_{k \neq i} | b_{ik} |,
\end{align*}
for $i=1,\ldots,n$.
It is shown in \cite{varah1975lower} that if $c_i(\bB) > 0$ and $d_i(\bB) > 0$ for all $i$, then
\[
\sigmin(\bB) \geq \min_{1 \leq i \leq n} \min\{ c_i(\bB), d_i(\bB)  \}.
\] 
We apply this result to $\tilde{\bR}_\setI^\setI$.
As considered previously, $\tilde{\bR}$ lies in $\setU^{N \times N}$.
This implies 
\[
\tilde{r}_{ii} = 1 - \sum_{k \neq i} \tilde{r}_{ki} \geq 1 - \frac{(N-1)}{4N} \geq \frac{3}{4},
\]
where the first inequality is due to $0 \leq \tilde{r}_{ki} \leq 1/(4N)$ for all $k \neq i$.
It follows that 
\[
c_i(\tilde{\bR}_\setI^\setI) \geq \frac{3}{4} - \frac{| \setI|}{4N} \geq \frac{1}{2},
\]
and that, similarly, $d_i(\tilde{\bR}_\setI^\setI) \geq 1/2$.
Hence we have $\sigmin(\tilde{\bR}_\setI^\setI) \geq 1/2$, and the proof is done.

\fi

\bibliographystyle{IEEEbib}
\bibliography{refs}

\newpage

\end{document}